# A Non-Orthogonal Distributed Space-Time Coded Protocol
# Part I: Signal Model and Design Criteria

G.Susinder Rajan and B.Sundar Rajan
Department of Electrical Communication Engineering
Indian Institute of Science, Bangalore, India
Email: {susinder, bsrajan}@ece.iisc.ernet.in

*Abstract*— **In this two-part series of papers, a generalized non-orthogonal amplify and forward (GNAF) protocol which generalizes several known cooperative diversity protocols is proposed. Transmission in the GNAF protocol comprises of two phases - the broadcast phase and the cooperation phase. In the broadcast phase, the source broadcasts its information to the relays as well as the destination. In the cooperation phase, the source and the relays together transmit a space-time code in a distributed fashion. The GNAF protocol relaxes the constraints imposed by the protocol of Jing and Hassibi on the code structure. In Part-I of this paper, a code design criteria is obtained and it is shown that the GNAF protocol is delay efficient and coding gain efficient as well. Moreover GNAF protocol enables the use of sphere decoders at the destination with a non-exponential Maximum likelihood (ML) decoding complexity. In Part-II, several low decoding complexity code constructions are studied and a lower bound on the Diversity-Multiplexing Gain tradeoff of the GNAF protocol is obtained.**

## I. INTRODUCTION & BACKGROUND

Recently there has been a growing interest in *cooperative diversity techniques*, wherein multiple terminals cooperate to form a virtual antenna array to leverage the spatial diversity benefits even if a local antenna array is not available. Since the works of [1]-[4], several cooperative transmission protocols have been proposed [5]-[10]. These protocols widely fall under two classes- Amplify and forward (AF) and Decode and forward (DF). The AF protocol is particularly attractive for two reasons - first, the operations at the relay nodes are considerably simplified, and second, we can avoid imposing bottlenecks on the rate by not requiring the relay nodes to decode [6]. Hence in this paper, we are mainly interested in the AF protocols although extension to the DF protocols can also be done.

The contributions of this paper are

- Propose a new AF protocol for cooperative diversity that generalizes the protocols of Jing and Hassibi [6], Nabar et. al. [7] and Azarian et. al. [5].
- It is shown using pair-wise error probability (PEP) analysis that a diversity order of $R+1$ is possible with $R$ relays, with the duration of the cooperation phase equal to $R$ channel uses whereas the known protocols [9] need at least $R+1$ channel uses.
- It is shown that any square complex orthogonal design achieves full diversity when employed in GNAF protocols.
- For the GNAF protocol, irrespective of the Distributed Space-Time Code (DSTC) used in the cooperation phase, the average sphere decoder complexity is not exponential even if the destination has only one receive antenna. This was not the case in some of the earlier proposed protocols [9].

The paper is organized as follows: In Section II, we describe the GNAF protocol and in Section III, we derive a code-design criteria based on PEP analysis. Also, we point out the delay optimality, coding gain advantage and sphere decoding complexity advantage. Simulation results are provided in Section IV.

The proof for all the theorems, lemmas and claims are omitted due to lack of space.

*Notation:* For a complex matrix $A$, $A^*$, $A^T$ and $A^H$ denote the conjugate, transpose and conjugate transpose respectively. $A_I$ denotes the real matrix obtained by taking the real parts of all the entries of the matrix $A$ and $A_Q$ denotes the real matrix obtained by taking the imaginary parts of all the entries of the matrix $A$. For a square matrix $B$, $|B|$ and $\text{Tr}(B)$ denote the determinant and trace of the matrix $B$ respectively.

## II. GNAF PROTOCOL DESCRIPTION

We consider a wireless relay network with one source, $R$ relays and a single destination $D$. The links between the different terminals are assumed to be fading links. The channel path gains from the source to the $i^{th}$ relay, denoted by $f_i$ and those from the $j^{th}$ relay to the destination denoted by $g_j$ are all assumed to be i.i.d $\mathcal{CN}(0,1)$. The channel path gain, $g_0$ from the source to the destination is also assumed to be $\mathcal{CN}(0,1)$. We assume that all the terminals are synchronized at the symbol level. Every transmission from the source to the destination comprises of two phases, i.e., the broadcast phase comprising of $T_1$ channel uses and the cooperation phase comprising of $T_2$ channel uses [1]. We next describe three different versions of the GNAF protocol.

**GNAF-I:** In the broadcast phase, the source terminal communicates with the relay and destination terminals. In the cooperation phase both the relays and the source communicate with the destination terminal.

**GNAF-II:** The broadcast phase is same as GNAF-I protocol. In the cooperation phase, only the relays communicate with the destination terminal.

**GNAF-III:** In the broadcast phase the source terminal communicates with the relays only. In the cooperation phase the source terminal as well as the relays communicate with the destination terminal.

In GNAF protocols, the relays are allowed to transmit only a complex linear combination of the symbols (received from the source terminal during the broadcast phase) as well as their conjugates. To perform these linear operations each relay is equipped with a pair of matrices called 'relay matrix pair'.

The signal model for GNAF-I protocol is as shown in (1) at the top of the next page, where

- $s$ is the vector transmitted by the source from a codebook consisting of $\mathscr{C} = \{s_1, s_2, \ldots, s_L\}$ complex vectors of size $T_1 \times 1$ satisfying $\mathrm{E}\left\{s^H s\right\} = 1$.
- $r_i$ denotes the received vector at the $i^{th}$ relay and $t_i$ denotes the vector transmitted by the $i^{th}$ relay. $A_i$ and $B_i$ are complex matrices of size $T_2 \times T_1$ satisfying $\parallel A_i \parallel_F^2 + \parallel B_i \parallel_F^2 \leq 1$. The pair of matrices $A_i$ and $B_i$ will be called the 'relay matrix pair' for the $i^{th}$ relay.
- $y_{D,1}$ and $y_{D,2}$ denote the received vector at the destination during the broadcast phase and cooperation phase respectively. $w_1$ and $w_2$ represent the additive noise at the destination whose entries are i.i.d $\mathcal{CN}(0,1)$. The quantities $\pi_1, \pi_2$ and $\pi_3$ are the power allocation factors satisfying $\pi_1 + \pi_2 + R\pi_3 = T_1 + T_2$ so that $P$ represents the total average power spent by the source and the relays together.

The received vector at the destination can be written

[1] The channel fade coefficients are assumed to remain constant for the entire duration (cooperation frame) of $T_1 + T_2$ channel uses and may vary independently from one cooperation frame to another.

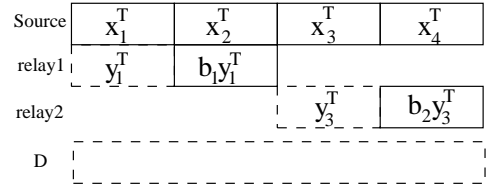

(a) NAF protocol

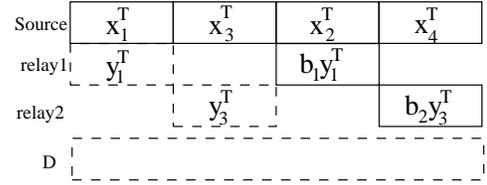

(b) GNAF–I protocol

Fig. 1. Frame structure of NAF and special case of GNAF-I for a two relay network

as follows

$$y = \left[ \begin{array}{c} y_{D,1} \\ y_{D,2} \end{array} \right] = \sqrt{\frac{\pi_3 \pi_1 P^2}{\pi_1 P + 1}} SH + W \qquad (2)$$

where $S, H$ and $W$ are as shown in (3) at the top of the next page. The DSTC in this case is the collection of all the $(T_1 + T_2) \times 2(R+1)$ matrices $S$ and we call this DSTC as GNAF-DSTC.

**The NAF protocol and GNAF-I protocol:** The NAF protocol proposed in [5] can be viewed as a special case of GNAF-I protocol. The frame structure of the NAF protocol for a two relay network is shown in Fig.1(a). Here $b_1$ and $b_2$ represent the scaling factor used by the relays in order to meet their respective power constraints. By simply permuting the time slots of the NAF protocol, we get the frame structure shown in Fig.1(b). It is easy to observe that the frame structure in Fig.1(b) is a special case of the GNAF-I protocol with the relay matrix pairs being $A_1 = \left[ \begin{array}{cc} b_1 & 0 \\ 0 & 0 \end{array} \right]$; $B_1 = 0$; $A_2 = \left[ \begin{array}{cc} 0 & 0 \\ 0 & b_2 \end{array} \right]$ and $B_2 = 0$. Moreover, since the performance of a ML receiver is invariant to permutations of the received vector, it follows that *the optimal DM-G tradeoff of the GNAF-I protocol is at least as good as the NAF protocol* [5].

### III. CODE DESIGN CRITERIA FOR GNAF-DSTCS

To simplify the PEP analysis, we will require the noise components of $n_i = A_i v_i + B_i v_i^*$ to be uncorrelated. Simplifying the expression for $n_i$ we get

$$\left[ \begin{array}{c} n_{iI} \\ n_{iQ} \end{array} \right] = \left[ \begin{array}{cc} A_{iI} + B_{iI} & -A_{iQ} + B_{iQ} \\ A_{iQ} + B_{iQ} & A_{iI} - B_{iI} \end{array} \right] \left[ \begin{array}{c} v_{iI} \\ v_{iQ} \end{array} \right].$$

$$\begin{align}
y_{D,1} &= \sqrt{\pi_1 P} g_0 s + w_1 \\
r_i &= \sqrt{\pi_1 P} f_i s + v_i, \forall \ i = 1, \ldots, R \\
t_i &= \sqrt{\frac{\pi_3 P}{\pi_1 P + 1}} (A_i r_i + B_i r_i^*), \forall \ i = 1, \ldots, R \\
y_{D,2} &= \sum_{i=1}^{R} g_i t_i + \sqrt{\pi_2 P} g_0 (A_0 s + B_0 s^*) + w_2
\end{align}$$
(1)

$$\begin{align}
S &= \begin{bmatrix} \sqrt{\frac{\pi_1 P+1}{\pi_3 P}} I_{T_1} s & 0 & \ldots & 0 & 0 & 0 & \ldots & 0 \\ \sqrt{\frac{\pi_2(\pi_1 P+1)}{\pi_3 \pi_1 P}} A_0 s & A_1 s & \ldots & A_R s & \sqrt{\frac{\pi_2(\pi_1 P+1)}{\pi_3 \pi_1 P}} B_0 s^* & B_1 s^* & \ldots & B_R s^* \end{bmatrix} \\
H^T &= \begin{bmatrix} g_0 & g_1 f_1 & \ldots & g_R f_R & g_0 & g_1 f_1^* & \ldots & g_R f_R^* \end{bmatrix} \\
W &= \begin{bmatrix} w_1 \\ \sqrt{\frac{\pi_3 P}{\pi_1 P+1}} \sum_{i=1}^{R} g_i (A_i v_i + B_i v_i^*) + w_2 \end{bmatrix}
\end{align}$$
(3)

From the above equation, it is clear that the required conditions for the noise components of $n_i$ to remain uncorrelated is that the matrix $\Xi_i = \begin{bmatrix} A_{iI} + B_{iI} & -A_{iQ} + B_{iQ} \\ A_{iQ} + B_{iQ} & A_{iI} - B_{iI} \end{bmatrix}$ should have the property that $\Xi_i \Xi_i^T$ is a diagonal matrix $\forall \ i = 1, \ldots, R$.

*Theorem 1:* Let $S_i$ be the transmitted codeword and $S_j$ be some other codeword. Let $\Delta S = S_i - S_j$ and

$$D = \begin{bmatrix} I_{T_1} & \\ & \left(1 + \frac{\mu \pi_3 P R}{\pi_1 P + 1}\right) I_{T_2} \end{bmatrix}$$

where, $\mu$ is the maximum variance of the noise components of $n_i = A_i v_i + B_i v_i^*$ over all $i = 1, \ldots, R$. If $\Xi_i \Xi_i^T$ is a diagonal matrix $\forall \ i = 1, \ldots, R$ and if $M = (\Delta S)^H D^{-1} (\Delta S)$ has rank $\leq (R+1)$, then for large $R$ and large $P$, the pairwise error probability (PEP) that a ML receiver erroneously decodes to $S_j$ can be upper bounded as

$$PEP \lesssim \left(\frac{\pi_3 \sigma_{min}^2}{4}\right)^{-\text{rank}(M)} P^{-\left(1 + (\text{rank}(M) - 1)(1 - \frac{\log(\log P)}{\log P})\right)}$$
(4)

where, $\sigma_{min}^2$ is the minimum non-zero eigen value of $M$.

*Proof:* Let $F = \begin{bmatrix} 1 & 0 & \ldots & 0 \\ 0 & f_1 & \ddots & \vdots \\ \vdots & \ddots & \ddots & 0 \\ 0 & \ldots & 0 & f_R \\ 1 & 0 & \ldots & 0 \\ 0 & f_1^* & \ddots & \vdots \\ \vdots & \ddots & \ddots & 0 \\ 0 & \ldots & 0 & f_R^* \end{bmatrix}$. For large $R$ and $P$, the PEP can be shown to be upper bounded by

$$PEP \lesssim \underset{\{f_i\}}{E} \left| I_{2(R+1)} + \frac{\pi_3}{4} P M F F^H \right|^{-1}.$$

Given that $\text{rank}(M) \leq (R+1)$, for large $P$, it can be shown that

$$PEP \lesssim \left(\frac{\pi_3 \sigma_{min}^2}{4}\right)^{-\text{rank}(M)} P^{-\left(1 + (\text{rank} M - 1)(1 - \frac{log(logP)}{logP})\right)}$$

■

Hence the design criteria is to maximize the rank of $\Delta S$. If we replace the identity matrix in the first column of $\Delta S$ by the all zero matrix, then it will correspond to GNAF-III protocol. It is easy to see that a diversity order of $R+1$ is possible in the GNAF-III protocol if $T_2 \geq (R+1)$. By setting $T_1 = T_2$, $\pi_2 = 0$ and replacing the identity matrix in the first column of $\Delta S$ by the all zero matrix, we get the Jing and Hassibi protocol [6]. Then, if $T_2 \geq R$, a diversity order of $R$ can be obtained. In [6], $A_i$ and $B_i$ were *restricted to real square matrices* and a code design criteria was obtained only for a more restrictive case, i.e., for any $i = 1, \ldots, R$, either $A_i = 0$ or $B_i = 0$. For the general case, the authors conjectured that if $T_2 \geq R$, a diversity order of $R$ was possible. This conjecture has been proved here for a much more general case assuming that the number of relays is large. Further, the design criteria to achieve the required diversity order has also been obtained.

Notice that the first column of $\Delta S$ is not in the linear span of the other columns of $\Delta S$. Hence it is sufficient for the sub matrix obtained by deleting the fist column and first $T_1$ rows of $\Delta S$ to satisfy full rank condition. Further note that even if we delete in addition the $(R+2)^{th}$ column of $\Delta S$ containing the matrix $B_0$, the maximum achievable diversity order is not disturbed. We call the resultant sub-matrix the 'extended relay matrix' which is shown below.

$$S_{ER} = \begin{bmatrix} A_1 s & \ldots & A_R s & B_1 s^* & \ldots & B_R s^* \end{bmatrix}$$

Hence if $T_2 \geq R$, a diversity order of $R+1$ can be achieved by GNAF-I as well as GNAF-II protocols. Thus $T_2 = R$ is the least possible delay. Notice an important fact that the *source continuing transmission even in the cooperation phase is not mandatory for getting diversity $R + 1$*. However, it controls the coding gain as will be shown later in this section.

The maximum diversity achievable by the protocol of [6] is only $R$. Therefore the new protocol increases the diversity order by one for the same number of relays. In [9], by some other modifications to the protocol in [6], a diversity order of $R+1$ is obtained. The GNAF protocol is differs from that of [9], since here the destination processes a vector of size $T_1 + T_2$ which is in contrast to processing a vector of size $T_2$ as assumed in [6], [8], [9]. Also the protocol in [8], [9] required $T_2 \geq (R+1)$ in order to achieve diversity $R+1$, whereas the GNAF-I,II protocols require only $T_2 \geq R$. Hence the GNAF protocol is *delay efficient*.

To highlight another advantage, let us consider $R+1$ relays for the protocol in [6] and $R$ relays for the GNAF-I,II and III protocols. Then the PEP expression derived in [6] for large $P$, is given by

$$PEP \lesssim cP^{-(R+1)\left(1-\frac{log(logP)}{logP}\right)} \quad (5)$$

where, $c$ is some positive constant. Note that for any $P$, the negative exponent of $P$ in the PEP (upper bound) expression (4) for the GNAF protocol is slightly more than the corresponding term in (5). Hence *the probability of error would decay faster* as a function of $P$ for the GNAF protocol. Intuitively, this happens because the transmission from the source to destination experiences only one channel fade coefficient whereas if the same transmission was made through the relay, then it would experience a product of two channel fade coefficients.

The generalized sphere decoding algorithm can be applied to decode $s$ from $y$. It can be observed that the size of vector $y$ is always greater than the size of vector $s$ for all practical choices of $T_1$ and $T_2$ in GNAF-I,II protocols. This leads to an equivalent channel model which is never under-determined. Hence the *a*verage sphere decoder complexity does not become exponential for any choice of $T_1$ and $T_2$ (or equivalently any GNAF-DSTC).

We will now give an example of a GNAF-DSTC which will achieve diversity $R + 1$ in GNAF-I,II protocols.

*Example 1:* Let $T_1 = T_2 = R = 4$. Consider the $4 \times 4$ complex orthogonal design which is shown below.

$$\Theta_{4(x_0,x_1,x_2)} = \begin{bmatrix} x_0 & x_1 & x_2 & 0 \\ -x_1^* & x_0^* & 0 & x_2 \\ -x_2^* & 0 & x_0^* & -x_1 \\ 0 & -x_2^* & x_1^* & x_0 \end{bmatrix}$$

Let us write it in the format required for us.

$$\Theta'_{4(x_0,x_1,x_2)} = \begin{bmatrix} x_0 & x_1 & x_2 & 0 & 0 & 0 & 0 & 0 \\ 0 & 0 & 0 & x_2 & -x_1^* & x_0^* & 0 & 0 \\ 0 & 0 & 0 & -x_1 & -x_2^* & 0 & x_0^* & 0 \\ 0 & 0 & 0 & x_0 & 0 & -x_2^* & x_1^* & 0 \end{bmatrix}$$

It is easy to check that $\Theta'_{4(x_0,x_1,x_2)} \Theta'^H_{4(x_0,x_1,x_2)} = (\sum_{i=0}^{2} |x_i|^2)I_4$. Thus the design $\Theta_{4(x_0,x_1,x_2)}$ satisfies the required rank criteria and hence can achieve diversity 5 in a 4 relay network. But note that only the rows of $\Theta'_{4(x_0,x_1,x_2)}$ are orthogonal but not the columns. Hence the most sought after *single symbol decodability feature of the complex orthogonal design is lost*.

In a similar manner, it can be easily shown that *any square complex orthogonal design achieves full diversity in GNAF protocols*. In [9], the use of orthogonal designs was suggested by doubling the number of symbols transmitted in the broadcast phase since complex conjugacy was not permitted at the relays in their protocol. However we see that if square orthogonal designs are employed, it is not actually necessary to do so.

### A. Coding Gain

In this subsection, we shall further consider only a special case among the class of linear dispersion codes wherein for any $i = 1, \ldots, R$, either $A_i = 0$ or $B_i = 0$ to show that GNAF protocol leads be higher coding gain compared to that of [6]. Without loss of generality, we shall assume that the first $N$ relays have $B_i = 0$ and the remaining $R-N$ relays have $A_i = 0$. Further we assume that $B_0 = 0$. Then the system model can be simplified to obtain

$$y = \begin{bmatrix} y_{D,1} \\ y_{D,2} \end{bmatrix} = \sqrt{\frac{\pi_3 \pi_1 P^2}{\pi_1 P + 1}} SH + W \quad (7)$$

where,

$$H^T = \begin{bmatrix} g_0 & \ldots & g_N f_N & g_{N+1} f_{N+1}^* & \ldots & g_R f_R^* \end{bmatrix} \quad (8)$$

and $S$ is given by the $S$ in (3) with appropriate columns dropped. If $M = (\Delta S)^H D^{-1}(\Delta S)$ is full rank for all pairs of distinct codewords, then it can be shown that

$$PEP \lesssim \left(\frac{4|M|^{\frac{1}{R+1}}}{\pi_3}\right)^{-(R+1)} P^{-\left(1+R\left(1-\frac{\log(\log P)}{\log P}\right)\right)}.$$

Thus $|M|$ can be taken to be a measure of coding gain at high SNR. For very small $P(P << 1)$ and large $R$, it can be shown that

$$PEP \lesssim \left(1 - \frac{\pi_3 \pi_1 P^2}{4} \text{Tr}(M')\right) + o(P^2) \quad (9)$$

where, $M' = (\Delta S)^H(\Delta S)$. Thus at low SNR, the design criteria is to maximize $\text{Tr}(M')$. Let $\Delta \hat{S}$ be as shown in (6) at the top of this page. Also, let $\hat{M} = (\Delta \hat{S})^H(\Delta \hat{S})$, $\Delta S_{ER} = \begin{bmatrix} A_1 \Delta s & \ldots & A_N \Delta s & B_{N+1} \Delta s^* & \ldots & B_R \Delta s^* \end{bmatrix}$ and let $M_{ER} = (\Delta S_{ER})^H(\Delta S_{ER})$. For large P, we

$$(\Delta \hat{S}) = \left[ \sqrt{\frac{\pi_2(\pi_1 P+1)}{\pi_3 \pi_1 P}} A_0 \Delta s \quad A_1 \Delta s \quad \ldots \quad A_N \Delta s \quad B_{N+1} \Delta s^* \quad \ldots \quad B_R \Delta s^* \right]. \quad (6)$$

have

$$|M| = \left( \frac{1}{1 + \frac{\mu \pi_3 R}{\pi_1}} \right)^{R+1} \left( |\hat{M}| + \left( \frac{\pi_1}{\pi_3} + \mu R \right) \| (\Delta s) \|^2 |M_{ER}| \right) \quad (10)$$

which shows that the *coding gain for GNAF-I protocol* ($|M|$) *is more* than that of GNAF-III protocol ($|\hat{M}|$) and Jing-Hassibi protocol ($|M_{ER}|$). Note that the coding gain for GNAF-I protocol ($|\hat{M}|$) depends upon the matrix $A_0$. Thus the power allocation factors $\pi_1, \pi_2, \pi_3$ and the matrix $A_0$ have to be carefully chosen to optimize the coding gain. Further we have $\mathrm{Tr}(M') = \mathrm{Tr}(\hat{M}) + \frac{1}{\pi_3 P} \| (\Delta s) \|^2$. Hence at low SNR also, GNAF-I protocol performs better. This is because the transmission by the source in the broadcast phase has been made use of at the decoder, and the source has been permitted to transmit all the time.

## IV. SIMULATION RESULTS

In this section, we present simulation results of the CIOD for $4$ relays to show that the GNAF-II protocol gives better error performance at all values of SNR compared to the Jing and Hassibi protocol [6]. The chosen values for the various parameters are $T_1 = T_2 = R = 4$, $\pi_1 = \frac{T_1+T_2}{2}$, $\pi_2 = 0$ and $\pi_3 = \frac{T_1+T_2}{2R}$. The signal set was taken to be QPSK rotated by $31.7175°$. Fig.2 shows the ML decoder's symbol error rate and codeword error rate performance comparison for both protocols. Observe that performance in the GNAF-II protocol uniformly dominates the performance in Jing and Hassibi protocol. Also observe that the probability of error decays faster in the GNAF-II protocol than in the Jing and Hassibi protocol. This is because the GNAF-II protocol offers a diversity order of $5$ as compared to $4$ by the Jing and Hassibi protocol.

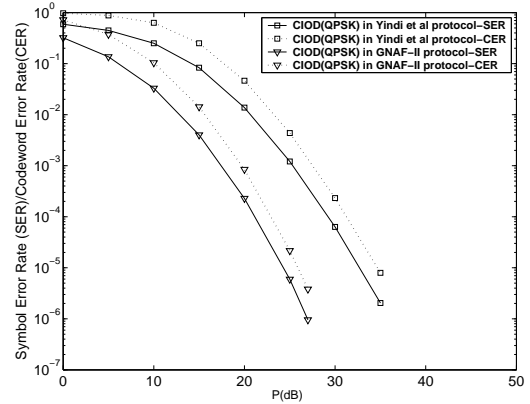

Fig. 2. Performance of CIOD for 4 relays in GNAF-II protocol and Jing and Hassibi protocol

## ACKNOWLEDGMENT

This work was partly supported by the DRDO-IISc Program on Advanced Research in Mathematical Engineering, partly by the Council of Scientific & Industrial Research (CSIR), India, through Research Grant (22(0365)/04/EMR-II) and also by Beceem Communications Pvt. Ltd., Bangalore to B.S. Rajan.